\documentclass[showpacs,epsfig,preprintnumbers,amsmath,amssymb,nofootinbib,20pt]{revtex4-1}
\pdfoutput=1
\usepackage{graphicx}
\usepackage{dcolumn}
\usepackage{bm}
\newcommand {\be}{\begin{equation}}
\newcommand {\ee}{\end{equation}}
\begin{document}
\begin{center}
\textbf{Dark Matter Decaying into Millicharged Particles as a Solution  to AMS-02 Positron Excess}
\end{center}
\newcommand{\AHEP}{%
School of physics, Institute for Research in Fundamental Sciences
(IPM)\\P.O.Box 19395-5531, Tehran, Iran\\

  }
\newcommand{\Tehran}{%
School of physics, Institute for Research in Fundamental Sciences (IPM)
\\
P.O.Box 19395-5531, Tehran, Iran}
\def\roughly#1{\mathrel{\raise.3ex\hbox{$#1$\kern-.75em
      \lower1ex\hbox{$\sim$}}}} \def\lsim{\roughly<}
\def\gsim{\roughly>}
\def\ltap{\raisebox{-.4ex}{\rlap{$\sim$}} \raisebox{.4ex}{$<$}}
\def\gtap{\raisebox{-.4ex}{\rlap{$\sim$}} \raisebox{.4ex}{$>$}}
\def\lsim{\raise0.3ex\hbox{$\;<$\kern-0.75em\raise-1.1ex\hbox{$\sim\;$}}}
\def\gsim{\raise0.3ex\hbox{$\;>$\kern-0.75em\raise-1.1ex\hbox{$\sim\;$}}}



\date{\today}
\author{Y. Farzan}\email{yasaman@theory.ipm.ac.ir}
\author{M. Rajaee}\email{meshkat.rajaee@ipm.ir}
\affiliation{\Tehran}
\begin{abstract}
	The positron excess observed by PAMELA and then confirmed by AMS-02 has intrigued the particle physics community since 2008. Various dark matter decay and annihilation models have been built to explain the excess. However, the  bounds from isotropic gamma ray  disfavor the canonical dark matter  decay scenario.
	We propose a solution to this excess based on the  decay of dark matter particles into intermediate millicharged particles which can be trapped by the galactic magnetic field. The subsequent decay of the millicharged particles to electron positron in our vicinity can explain the excess. Since these particles diffuse out of the halo before decay, their contribution to the isotropic gamma ray background is expected to be  much smaller than that in the canonic dark matter decay scenarios.
We show that the model is testable by direct dark matter search experiments.	
\end{abstract}
\date{\today}
\maketitle

\section{Introduction}
 In the standard cosmic ray model,  positrons are produced by  inelastic scattering of primary cosmic rays (mainly protons) off interstellar matter (i.e., hydrogen)  \cite{Nagano:2000ve}. In 2008, Payload for Antimatter Matter Exploration
and Light-nuclei Astrophysics (PAMELA) discovered an excess of the positron-to-electron ratio above $\sim$10 GeV \cite{Adriani:2008zr}, confirming the previously reported excess by HEAT \cite{Barwick:1997ig}. 
In subsequent years,  the AMS-02  experiment \cite{Aguilar:2013qda}  with a more sensitive detector and a wider energy sensitivity range confirmed the excess. 

Origin of the excess positrons is  not certain.  Contributions from pulsars \cite{Hooper:2008kg}, secondary cosmic ray from supernova remnants \cite{DiMauro:2014iia,Fujita:2009wk,Kohri:2015mga,Dunsky:2018mqs} and decaying or annihilating Dark Matter (DM) \cite{ArkaniHamed:2008qn,Belotsky:2014nba,Belotsky:2014haa} are among possible explanations suggested in the literature.
None of these solutions has been completely established as the prime origin. For example, although the recent observations by HAWC on Geminga and Monogem confirm that the energetic of the positron signal from close-by pulsars can match the observed positron excess \cite{Hooper:2017gtd},  the diffusion parameters derived from observation do not seem to be compatible with a pulsar solution to the AMS-02 positron excess \cite{Abeysekara:2017old}. In this paper, we will focus on the possibility that dark matter is responsible  for  all the positron excess.

 Models of annihilating DM  need large enhancement on the annihilation cross section \cite{Yuan:2013eja}, which can occur via  the Sommerfeld enhancement \cite{ArkaniHamed:2008qn}
or by the Breit-Wigner enhancement \cite{Guo:2009aj}. These models lead to delayed recombination problem so they are disfavored \cite{Slatyer:2009yq}. 
 More recently dark matter annihilation solution is further constrained by the bound on new sources of energy injection during dark ages by the EDGES data on the 21 cm absorption line \cite{Liu:2018uzy}.
 An independent constraint comes from the limit on diffuse $\gamma$-ray observed by Fermi-LAT. The electron and positron produced by DM  can go through inverse Compton scattering on CMB, giving rise to a gamma ray flux. This bound disfavors the DM decay solution to the positron excess  \cite{Blanco:2018esa,Cirelli:2012ut}. 
 Within the decay scenario, the signal from a volume of DM is proportional to $\int_V \rho_{DM}/|\vec{r}|^2 dV$. The contributions from dark matter decay inside the halo and from the extragalactic DM turn out to be comparable  \cite{Cirelli:2012ut}. That is while $\rho_{DM}/|\vec{r}|^2$ inside the halo is much larger than that outside the halo, the volume outside is much larger. Instead of prompt decay into $e^- e^+$, if DM particles decay into meta-stable particles that diffuse out of the halo before decay, the bound can  therefore significantly relax \cite{Kim:2017qaw}. This is the basis of the idea proposed in this paper.
 
 On one hand, we want the intermediate particles produced in the halo to go out of the halo before decay and on the other hand, we want those produced in the disk to remain in our vicinity.
 Similar to the idea proposed in \cite{meshkat}, this can be achieved by millicharged intermediate particles. Such particles become trapped by galactic magnetic field but they 
 can escape the halo (where the magnetic filed is small) with a speed close to that of light. 
 
 This paper is organized as follow:
 In section \ref{positron excess}, we  introduce our solution to the AMS-02 positron excess. Moreover, we discuss various bounds on the parameters of our scenario and the prospect of testing it by future  experiments. In section \ref{model}, we introduce an underlying  model embedding the scenario. Finally, section \ref{Summary} is devoted to summary and discussion.

\section{A solution to positron excess}\label{positron excess}

\begin{figure}
	\begin{center}
		\includegraphics[scale=0.4]{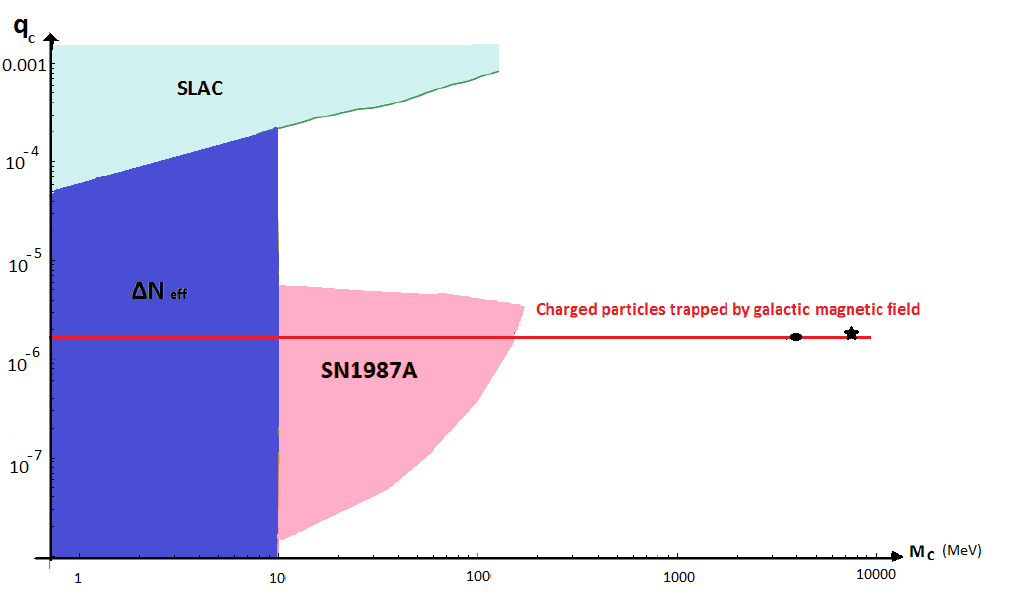}
	\end{center}
	\caption{Bounds on the charge of $C$-particle versus its mass. Light blue region is excluded by the SLAC experiment \cite{Prinz:1998ua}. The dark blue region is excluded by  the BBN constraint \cite{Boehm:2013jpa} and the pink region is excluded  by supernova $1987A$ \cite{Chang:2018rso}. The horizontal line shows  lower limit on $q_C$ above which $C$ particles with energy $E_C=4$ TeV  have Larmour radius below $500$ pc  for galactic magnetic field of $B = 1 ~{\rm \mu G}$. The black dot and star indicate best fit point values to AMS-02 positron excess assuming EinastoB and NFW halo profiles as is shown in table 1.} \label{Fig:limit}
\end{figure}

We assume that dark matter consists of scalar meta-stable particles, $X$,  that can decay into millicharged $C \bar{C}$ particles with decay rate, $\Gamma_X$, much smaller than the inverse of the age of the universe.
For correctness and simplicity, we take the millicharged particles in this scenario scalar but similar argument holds valid with fermionic millicharged particles.  
 Similarly to other dark matter scenarios designed to explain the high energy cosmic positron excess, the $X$ particles should be  heavier than TeV.   Because of the small but nonzero electric charge of the $C$ and $\bar{C}$ particles,  the magnetic field in galaxy can  keep the $C$ and $\bar{C}$ particles inside  the disk. 
For this purpose, the Larmour radius in the typical  interstellar magnetic field should be much smaller than the galactic disk thickness.
Since the dark matter particles in the galaxy are non-relativistic, the energy of $C$ and $\bar{C}$, $E_C$, produced from the $X$ decay will be equal.
 Taking $m_C\ll m_X$, the momentum of the $C$ and $\bar{C}$ particles at production will be $p_C\simeq E_C\simeq m_X/2$ so the Larmour radius can be estimated as $r_L=E_C/(q_C B)$. Taking $r_L\sim 500$ pc, $B\sim \mu$G \cite{Sun:2010sm} and $E_C\sim 4$ TeV, we find that the electric charge of the $C$ particles has to be given by
\begin{equation} q_C \sim 1.5 \times 10^{-6} \frac{500~{\rm pc}}{r_L} \frac{E_C}{4~{\rm TeV}}\frac{\mu {\rm G}}{B}  \ . \end{equation}
As seen from Fig. \ref{Fig:limit}, while the bound from SLAC is too weak to be limiting for our scenario, the BBN bounds set a lower bound of 10 MeV on $m_C$. We shall scrutinize the bounds from early universe more thoroughly later in this section. 
Moreover to be safe from the most conservative supernova bounds, the $C$ particles should be heavier than 100 MeV. In order to explain the positron excess we assume the existence of another millicharged particle denoted by $C'$ with the same electric charge and with an effective coupling of 
\be\frac{ C'^\dagger C \bar{e}e}{\Lambda_C}. \label{cc'}\ee
In the next section, we shall introduce the underlying model that gives rise to this effective interaction.
 The  differential decay width in the rest frame of $C$ is then given by
 \begin{equation} \label{diff}
\frac{d \Gamma (C \rightarrow e^- e^+ C^\prime) }{d E_e} = \frac{  E_e ^2}{64 \pi^3 m_C ^2 \Lambda _C ^2 } \frac{(2 m_C E_e - m_C ^2 + m_{C^\prime} ^2)^2}{(m_C -2E_e)^2},
\end{equation}
neglecting $m_{C^{\prime}}$, the total decay width can be written as
\begin{equation} 
\Gamma (C \rightarrow e^- e^+ C^\prime) = \frac{ m_C ^3 }{1536 \pi ^3 \Lambda_C ^2}.
\end{equation}

In order to soften the bound from non-observation of the gamma ray signal from halo, the lifetime of the $C$ particles has to be long enough to escape the halo: 
\begin{equation}
\frac{m_C}{E_C}\Gamma (C \rightarrow e^- e^+ C^\prime)< \frac{1}{5} \times 10^{-5} ~{\rm yr^{-1}}.
\end{equation}
On the other hand unless the $C$ particles decay faster, supernova shock waves can pump energy to the $C$ particles
driving them out of galaxy disk within a time scale of 100 Myr \cite{Chuzhoy:2008zy,Dunsky:2018mqs}. We therefore assume that
\begin{equation}
\frac{m_C}{E_C}\Gamma (C \rightarrow e^- e^+ C^\prime) > 10^{-7} ~{\rm yr^{-1}},
\end{equation}
which means the decay takes place before supernova shock waves can significantly accelerate the $C$ particles.\footnote{Notice however that if the lifetime (in the galaxy frame) is between $10^7-10^8$ years, the $C$ particles can obtain significant energy from supernova shock waves before decay, opening the possibility that lighter dark matter particles ($X$ particles)
	also explain the positron excess. We  shall not however explore this possibility in the present paper.}
Thus, we find
\be  5 \times 10^{15}~{\rm GeV} \left( \frac{8~{\rm TeV}}{m_X}\right)^{1/2}\left( \frac{m_C}{4~{\rm GeV}}\right)^2
<\Lambda_C< 1.5  \times  10^{17}~{\rm GeV} \left( \frac{8~{\rm TeV}}{m_X}\right)^{1/2}\left( \frac{m_C}{4~{\rm GeV}}\right)^2. \label{range} \ee
The $C'$ particles  which are stable will be eventually driven out by the supernova shock waves.
Notice that the magnetic field in the galaxy has an axial symmetry \cite{Jansson:2012pc}. The $C$ particles will spiral around
the magnetic fields which themselves circle around the galaxy center. Thus, the $C$ particle decaying in our vicinity may have been produced in another part of the galaxy but still at distance of $r_\odot \pm r_L$ from the galaxy center (where $r_\odot \simeq 8$ kpc is the distance of the Sun from the galaxy center.) Thus, because of the spherical symmetry of the halo profile, the dark matter density at the $C$ production will be taken to be equal to that in our vicinity: $\rho_X=0.3-0.8$ GeV/cm$^3$ \cite{Benito:2019ngh}.

 The spiraling  
 $C$ and  $C^\prime$ particles will lose energy via synchrotron radiation  given by
\begin{equation}
\frac{dE_C}{dt} = -\frac{2}{3} ( \frac{E_C  }{m_C ^2} ) ^2 q_C ^4 B^2.
\end{equation}
The cooling time scale of $C$ particles  via synchrotron radiation  is much longer than  the time scale of the energy gain from  supernova shock waves  (100 Myr):
\begin{equation}
\frac{E_C}{ \mid \frac{dE_C}{dt} \mid } = 9.2 \times 10^{35} ~{\rm Myr} \times (\frac{m_C}{4 ~{\rm GeV}})^4 (\frac{1.5 \times 10^{-6}}{q_C})^4 (\frac{1~ \mu {\rm G}}{B})^2 (\frac{4 ~{\rm TeV}}{m_X}) \gg 100 ~{\rm Myr} ; 
\end{equation} 
Thus, the synchrotron energy loss is completely negligible.


\begin{table}[ht]
	\caption{Best fit point values to AMS-02 positron excess for different  assumptions on the positron energy loss function.} 
	\centering 
	\begin{tabular}{ccccc} 
		\hline\hline 
		DM halo Profile     &    $\chi^2$  & $m_C$ (GeV)  & $m_X$ (GeV)  & $\Gamma ~{\rm (sec^{-1})} $ \\ [2ex]
		\hline 
		
		NFW      &     56.52   & 8 &  10000 & $3.5 \times 10^{-27}$ \\ [2ex]
		\hline
		
		EinastoB     &     52.11 & 4 & 8000  & $2.7 \times 10^{-27}$  \\ [2ex]
		\hline
		

		\hline\hline 
		
	\end{tabular}
	\label{table} 
\end{table}

As is shown in \cite{Cirelli:2010xx}, the differential flux of positrons  from dark matter decay can be written as 
\begin{equation}
\frac{d \Phi _{e^+} (E  )}{d E} = \frac{1}{4 \pi b(E ) } (\frac{\rho_X}{m_X}) \Gamma (X \rightarrow C \bar{C}) \int _E ^{E_{max}}
d E_s \frac{dN (E_s)}{dE_s} \mathcal{ I} ( E , E_s )  
\end{equation}
where $E_s$ and $E$ are respectively positron energy at source and at detector. Notice that due to the energy loss, $E_s<E$ and maximum $E$ is equal to $E_{max}=m_X/2-m_e-m_{C'}$ which for $m_{C'},m_C\ll m_X$ can be approximately written as $E_{max}\simeq m_X/2$. $ b(E)= E^2   / (~{\rm GeV} \tau_\odot
 ) $ is the energy loss coefficient function with $\tau_\odot = 5.7 \times 10^{15} ~{ \rm sec}$ \cite{Cirelli:2010xx}. $dN(E_s)/dE_s$ gives the spectrum of positron at production from $C$ decay in the galaxy frame and is related to the differential 
 decay rate  at the $C$ rest frame (Eq. \ref{diff}) by a boost with $\gamma_C=m_X/(2 m_C)$. 
$I(E,E_s)$ is the  halo function that takes care of the energy loss of positrons in the galaxy before reaching the detector.
To carry out the analysis, we use the so-called reduced halo function for $I(E,E_s)$ with central values of parameters for EinastoB profile enumerated in Ref. \cite{Cirelli:2010xx}. We then check the robustness of our results against
different forms of $I(E,E_s)$ that are described in Ref. \cite{Cirelli:2010xx}. 

\begin{figure}
	\begin{center}
		\includegraphics[scale=0.45]{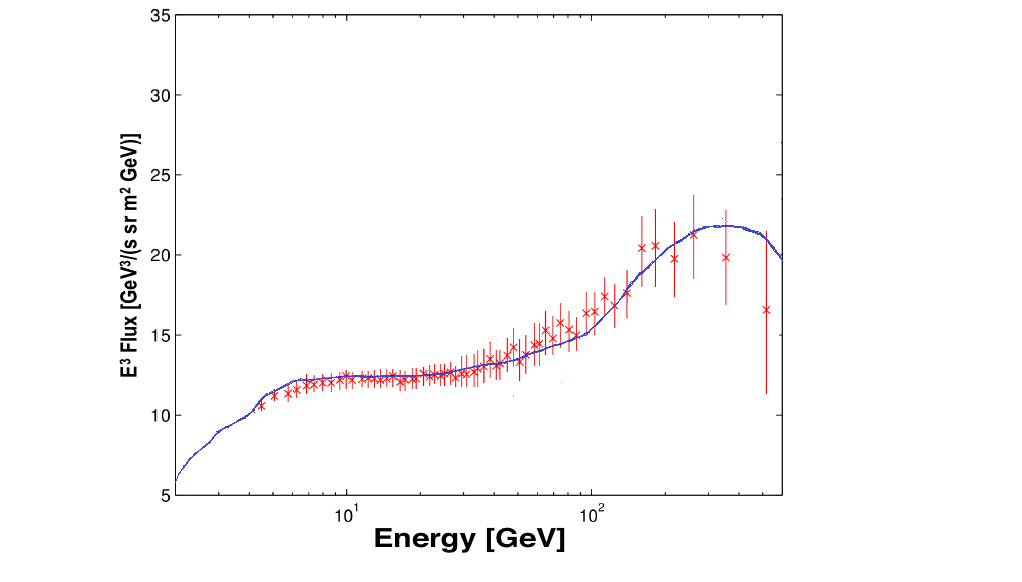}
	\end{center}
	
	\caption{The AMS-02 positron flux  compared with the prediction of our models.  The red dots represent the AMS-02 data with their experimental errors shown by the vertical bars \cite{Aguilar:2013qda}.  The  blue curve indicates expected positron spectrum plotted for  our best fit point of $ \Gamma_X= 2.7 \times 10^{-27} ~ {\rm sec^{-1}}, \ m_X=8  ~ {\rm TeV}  \  {\rm and} \ m_C=4 ~ {\rm GeV}$ plus cosmic ray positron background. }\label{fit}
	
\end{figure}

We should now find out what are the values of the parameters of the model that  explain the AMS-02 positron excess. For simplicity, we take $m_{C'}\ll m_C$. The exact value of $C'$ is not then  relevant for  the fit. However, even in the limit of $m_C\ll m_X$, the exact value of $m_C$ will affect the fit to the low energy part of the spectrum as the minimum $E_s$ at the galaxy frame is given by $m_C^2/(2m_X)$. We take $\Gamma_X$, $m_X$ and $m_C$ as free parameters to fit the data. We define $\chi^2$ as follows
\begin{equation} \label{chi2}
\chi^2=
\sum_{bins}\frac{[N^{pred} _i - N^{obs} _i]^2 }{\sigma_i^2}
\end{equation}
where $i$ runs over the energy bins. $N^{obs}_i$ is the observed number of events at each bin and $N^{pred}_i$ is the predicted number of events which is equal to the number of events from the $X$ decay in the ``$i$"th bin plus the cosmic ray background. We take $N^{obs}_i$ for positron flux of AMS-02 and the background from \cite{2016-AMS02-CERN}. The uncertainty in each bin, $\sigma_i$, comes from the uncertainty in the observed data  ($\sigma_i  (obs)$) as well as from the uncertainty in the background ($\sigma_i (bck)$) \cite{Tomassetti:2017hbe}: $\sigma_i=\sqrt{\sigma_i ^2 (obs) + \sigma_i ^2 (bck)  }$. The maximum bin energy is 580 GeV and we consider only the data points with energy above 3 GeV. Below this limit, the solar modulation  with large uncertainties are relevant \cite{Bindi:2017nee} which needs special treatment. For EinastoB dark matter profile, we find that the best fit can be achieved for
\begin{equation} \Gamma_X= 2.7 \times 10^{-27} ~ {\rm sec^{-1}}, \ m_X=8  ~ {\rm TeV}  \  {\rm and} \ m_C=4 ~ {\rm GeV} \label{best-fit}\end{equation}
with $\chi^2=52.2$ for  $64-3=61$ degrees of freedom and a p-value equal to 0.78.
The $N^{obs}$ and $N^{pred}$ for our fit are shown in Fig. \ref{fit}. We also redid the analysis for the energy loss function for the NFW dark matter profile with central values \cite{Cirelli:2010xx}.
The results are displayed in table 1. Comparing the two results, we deduce that although the goodness of fit remains excellent varying energy loss function but the values of the best fit parameters considerably change with the energy loss function. Notice that in our fit we have only considered  the AMS-02 positron excess data. 

Data on the $e^-+e^+$ flux from AMS-02 as well as from CALET \cite{Adriani:2018ktz} is also available. In order to check whether our best fit points are consistent with this data, we have also computed $\chi^2$ defined in Eq. (\ref{chi2}) for the $e^-+e^+$ spectrum. We have taken the $e^-+e^+$ background and its uncertainties  from \cite{Feng:2012gs}. Again because of the solar wind modulation, we have only included data points with energies above 3 GeV.  Data points include 69 points from AMS-02 taken from \cite{2016-AMS02-CERN} and 40 points from CALET taken from \cite{Adriani:2018ktz}.  Plugging in the best fit values shown in Eq. (\ref{best-fit}), we find $\chi^2=128.35$ which for $69+40=109$ degrees of freedom amounts to a p-value of 0.1 which is a reasonable goodness of  fit. We also searched for the best fit value for the $e^-+e^+$ flux from CALET and AMS-02  and found that 
 the best fit can be achieved for
\begin{equation} \Gamma_X= 10^{-27} ~ {\rm sec^{-1}}, \ m_X=5.5  ~ {\rm TeV}  \  {\rm and} \ m_C=4 ~ {\rm GeV} \label{best-fit-calet}\end{equation}
with $\chi^2=128$ for  $109-3=106$ degrees of freedom and a p-value equal to $0.071$  which indicates that they are consistent with each other. DAMPE \cite{Ambrosi:2017wek} and Fermi-LAT \cite{Loparco:2017spm} have also measured the $e^-+e^+$ flux. We do not however  include the DAMPE and Fermi-LAT data points which are slightly higher around 1 TeV. As discussed in \cite{Ambrosi:2017wek}, the discrepancy can be due to the uncertainty in the absolute energy scale. Notice that in our analysis, we have not included the energy uncertainty in $\sigma_i$. Allowing for this uncertainty, the acceptable range of parameters will further widen but exploring all these possibilities is beyond the scope of the present paper.

Since $X$ decay produces $e^-e^+$ after about $5 \times 10^5-10^7$ years, the recombination era as well as dark ages can be affected so we must check for the bounds from delayed recombination derived from CMB as well as
from the 21 cm bounds from EDGES. For this mass range, the strongest lower bound
on the dark matter lifetime is $10^{25}$ sec \cite{Liu:2018uzy} (see also \cite{Mambrini:2015sia})  so the values of $\Gamma_X$ that we have found (see table 1) are acceptable.

The millicharged particles $C$, $\bar{C}$,  $C'$ and $\bar{C'}$ in the early universe can be produced via Drell-Yan annihilation of SM fermions such as $e^-e^+\to C \bar{C}$ or $C' \bar{C'}$. With  $q_c \sim 10^{-6}$, the rates 
of $C\bar{C}$ and $C'\bar{C'}$ productions will be high enough to bring these particles  to thermal equilibrium. That means the stable $C^\prime$ particles (produced either directly or via $C$ decay) will contribute to dark matter. As shown in \cite{Dunsky:2018mqs}, from direct dark matter search experiments strong bounds can be set on the fraction of dark matter in the form of millicharged particles. To reduce the fraction below the bounds, a new annihilation mode for the $C\bar{C}$ and  $C'\bar{C'}$ pairs should open up. Within the mechanism that induces fractional electric charge to the $C$ and $C'$ particles such a mechanism can naturally emerge. The mechanism includes a new $U(1)$ gauge symmetry under which the $C$ and $C'$ particles are charged. As shown in \cite{Feldman:2008xs}, the kinetic mixing between this new $U(1)$ gauge boson and the hypercharge gauge boson leads to a tiny  electric charge for the new particles. In addition to the SM gauge bosons, there will be a new gauge boson which we denote by $\gamma'$. \footnote{ There 
 are three $s$-channel contributions to the $e^-e^+ \to C\bar{C}$ processes via the exchange of $\gamma$, $\gamma'$ and $Z$. For $m_{\gamma'}\ll m_C$, there can be partial cancellation between the
contributions from $\gamma$ and $\gamma'$ exchange such that the corresponding cross section is suppressed by $m_{\gamma'}^2/m_C^2$. Despite this suppression still $C\bar{C}$ particles can reach thermal equilibrium with the
plasma. The same argument holds valid for $C'$ and $\bar{C'}$, too.  
} Taking the new gauge coupling to be $g''$ and $m_{\gamma'}\ll m_C$, we can write
\begin{equation} \sigma (C\bar{C} \to \gamma'\gamma')\sim \frac{g''^4}{4\pi m_C^2}=1.87 \times 10^6 g''^4~{\rm pb} \left(\frac{4~{\rm GeV}}{m_C}\right)^2.\end{equation}
We can also write a similar formula for $C'\bar{C'} \to \gamma'\gamma'$. The fraction of dark matter in the form of ${C}'$ ($f_{C'}$) can be written
as $f_{C'}=5 \times 10^{-7} g''^{-4} (m_C ~{\rm or}~ m_{C'}/4~{\rm GeV})^2. $
Taking $g''\sim 1$, $f_{C'}$ will be low enough to satisfy the most stringent bounds from  direct dark matter search experiments  \cite{Dunsky:2018mqs}. Moreover, with such small $f_C$ and $f_{C'}$ at recombination era,
the energy dump from annihilation can be neglected. That is  $f_C^2 \sigma$ will be smaller than the bound from CMB \cite{Galli:2013dna}.
 
 The produced $\gamma'$ particles will decay into $e^-e^+$ with a rate of 
$$\Gamma_{\gamma'}\sim m_{\gamma'}\frac{g''^2 \delta^2}{4\pi}\sim ( 10^{-11}{\rm sec})^{-1}\left(\frac {g''\delta }{1.5 \times 10^{-6}}\right)^2 \frac{m_{\gamma'}}{200~{\rm MeV}}$$
 where we have taken the general case where the coupling of $\gamma'$ to the SM charged particles is of order of $g''\delta$ in which $\delta$ is the kinetic mixing and is of order of $q_C$. This means $\gamma'$ will decay into $e^-e^+$ long before the onset of the big bang nucleosynthesis era. \footnote{Notice that in the particular case when a certain relation between gauge boson mass mixing and kinetic mixing holds, the couplings of $\gamma'$ to SM fermions vanish \cite{Feldman:2007wj}, making $\gamma'$ stable. We do not, however, assume such relation. }

\section{The underlying model} \label{model}
\begin{figure}[t]
	\begin{center}
		\includegraphics[scale=0.4]{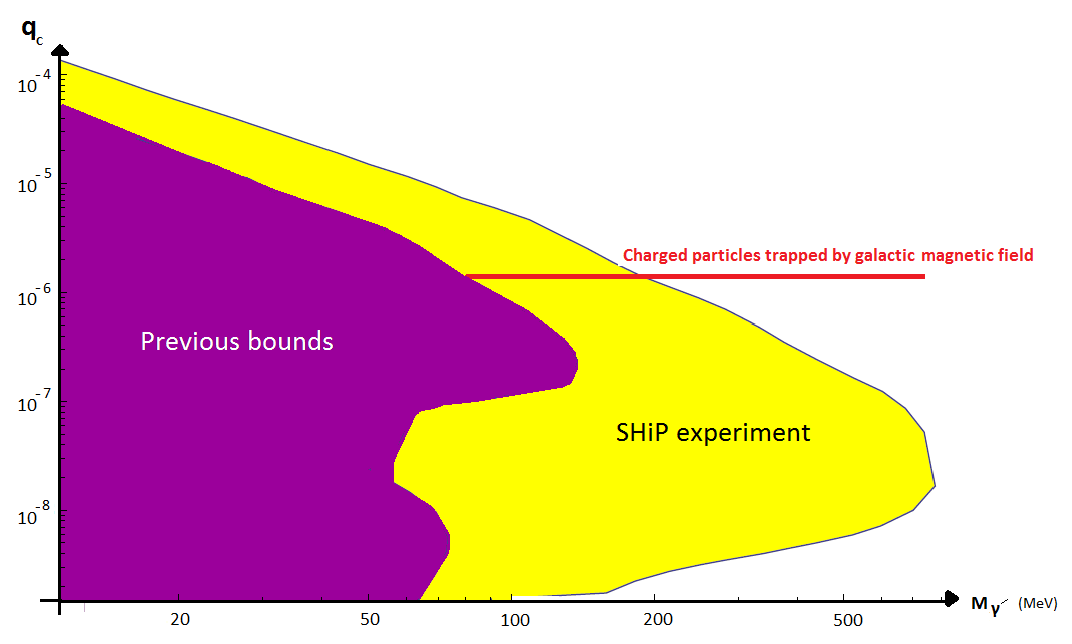}
	\end{center}
	
	\caption{ Beam dump experiment sensitivity contours for electric charge  of $ C$ particles  as a function of the $\gamma ^\prime$ mass. Purple region indicates excluded parameter space by previous experiments  \cite{Batell:2009di,Izaguirre:2013uxa,Batell:2014mga,Kahn:2014sra,Fradette:2014sza,Graverini:2015dka}. The yellow region shows the capability of the SHiP experiment to probe our model at $90 \%$ C.L.,  assuming a background of $0.1$ events
		for expected total exposure of $2 \times  10^{20}$ proton on target.}\label{bound}
\end{figure}

In this section, we elaborate on the underlying model  that gives rise to interaction forms required to realize the present scenario.  A central point to the scenario is the existence of  millicharged $C$ and $C'$ particles which for simplicity were taken to be scalars. Notice that in our scenario DM is electrically neutral and, unlike 
{\it e.g.} \cite{Foot:2014uba}, does not consist of millicharged particles. As mentioned before,
the $C$ and $C'$ particles can acquire tiny electric charges by adding a new $U(1)$ gauge symmetry under which they have the same charge. The details can be found in \cite{Feldman:2007wj} so we shall not repeat it here. For the range of electric charge of our interest,
the $\gamma'$ particle should be heavier than 80 MeV to avoid bounds from the present beam dump experiments \cite{Batell:2009di,Izaguirre:2013uxa,Batell:2014mga,Kahn:2014sra,Fradette:2014sza,Graverini:2015dka}. The upcoming SHiP experiment can probe the existence of $\gamma'$ corresponding to $q_C=1.5 \times 10^{-6}$ from $m_{\gamma'}=80$ MeV up to $m_{\gamma'}=200$ MeV (see Fig. \ref{bound}).
The SHiP experiment is a proposed fixed target experiment at the CERN with $400$ GeV proton beam \cite{Bonivento:2013jag}. 
For SHiP sensitivity predictions, a background of $0.1$ events for expected total exposure of $2 \times 10^{20}$ proton on target  is assumed \cite{Graverini:2015dka}.

\begin{figure}
	\begin{center}
		\includegraphics[scale=0.13]{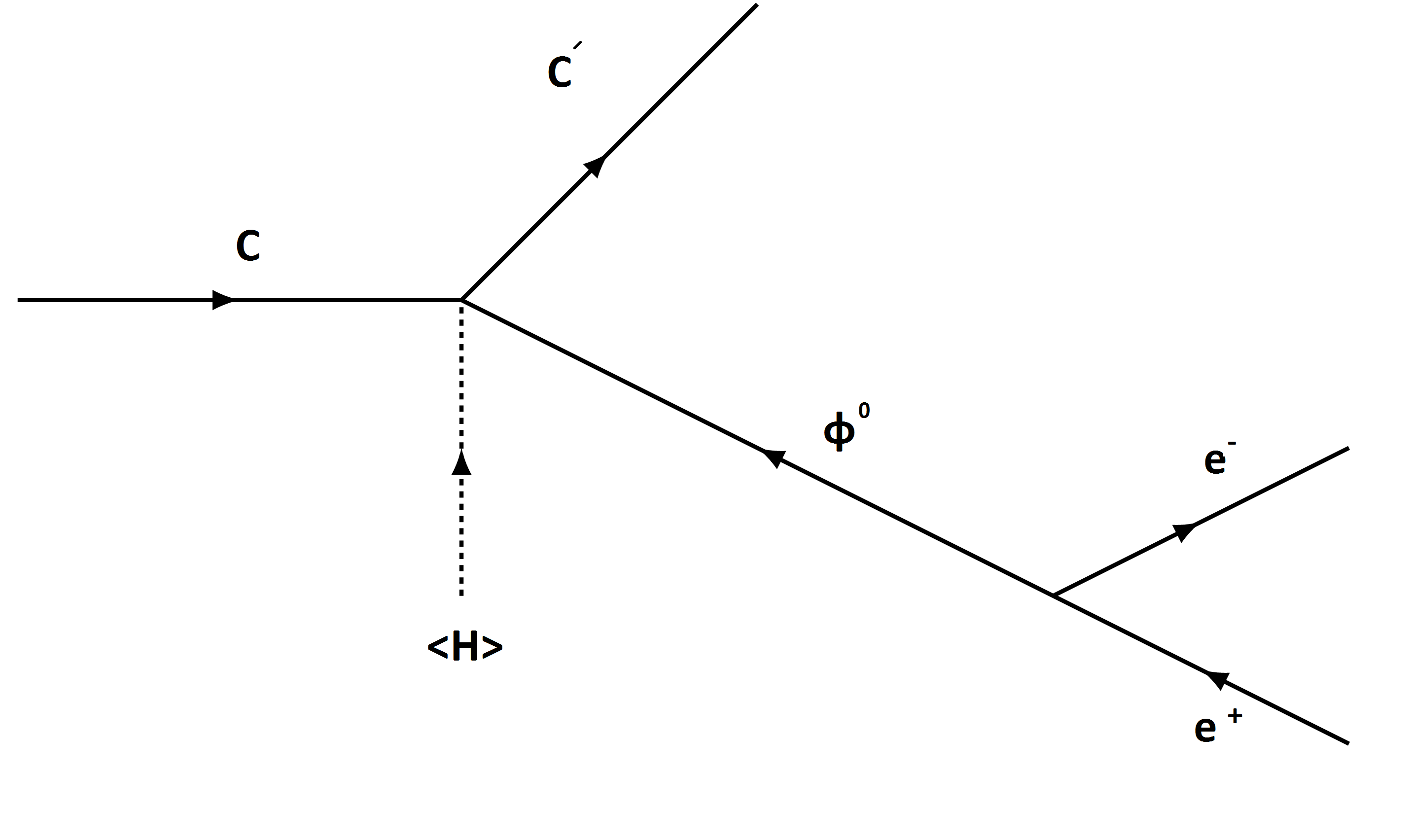}
	\end{center}
	\caption{$C$ particle  three body decay into $C^\prime$, $e^+$ and $e^-$}\label{decay}
\end{figure}

 Still we have to provide an underlying model for the effective action in Eq. (\ref{cc'}).
 An effective interaction of this type can be obtained by introducing a singlet scalar with a trilinear coupling to $\bar{C} C'$ and a Yukawa coupling to $e^-e^+$ through mixing with the SM Higgs. The $C$ and $C'$ will then  couple to also the other SM fermions with an effective coupling proportional to the mass of the fermions. 
 The $C'\to C\mu^-\mu^+$ and $C'\to Cq \bar{q}$ processes will then dominate over  $C'\to C e^-e^+$. To avoid these decay modes instead of introducing a singlet scalar, we introduce a doublet scalar with the same quantum numbers as those of the standard model Higgs, $\Phi_D^T=(\Phi^+, \Phi^0)$. Like the inert two Higgs doublet models \cite{Deshpande:1977rw,Barbieri:2006dq,LopezHonorez:2006gr}, we focus on a part of the parameter space where the new doublet does not develop a VEV.  For simplicity, let us introduce an approximate global $U_D(1)$ symmetry under which $\Phi_D \to e^{i \alpha_D}\Phi_D$ and $C'\to e^{i\alpha_D} C' $. The most general  potential involving the scalars can then be written as 
 $$V=V_H+V_\Phi+V_{H \Phi}+V_{\Phi C H}$$
 where $V_H$ is the standard Higgs potential,
 $$ V_\Phi=m_D^2 \Phi_D^\dagger \Phi_D+\frac{\lambda_D}{2}(\Phi_D^\dagger \Phi_D)^2$$
 and 
 $$V_{H\Phi}=\lambda_1 (\Phi^\dagger_D\Phi_D)(H^\dagger H)+\lambda_2 |\Phi_D^\dagger H|^2.$$
 Notice that a SU(2)$\times$U(1) invariant term of form $|\Phi_D^T \epsilon H|^2$ can be rewritten as a linear combination of the $\lambda_1$ 
 and $\lambda_2$ terms. The [$(\Phi_D^\dagger H)^2+H.c.$] term is forbidden by the global $U_D(1)$ symmetry so the real and imaginary components of $\Phi^0$ remain degenerate.
 Taking $m_D^2+(\lambda_1+\lambda_2)v^2/2>0$ and $\lambda_1, \lambda_D>0$, the minimum of the potential will remain at $\langle \Phi_D\rangle=0$.
 Finally, $V_{C\Phi H}$ contains all electroweak and $U_D(1)$ invariant renormalizable combination of $C$, $C'$, $H$ and $\Phi_D$. Notice that the mass mixing term $C^\dagger C'$ as well as quartic terms such as  $C^\dagger C'|H|^2$ and $C^\dagger C'|\Phi|^2$ are forbidden by $U_D(1)$. In the absence of this symmetry, $C^\dagger C' |H|^2$ along with the Higgs Yukawa couplings could lead to fast $C \to C' \bar{q}q ~{\rm or}~C' \bar{\mu}\mu $.  The $U_D(1)$ symmetry allows the following term
 $$\lambda_{CC'} (C')^{\dagger}C H^\dagger \Phi_D+H.c.$$
 We however impose an approximate $Z_2$ symmetry under which only $C'$ is odd. The $\lambda_{CC'}$ term breaks this symmetry and its smallness is explained by this approximate $Z_2$ symmetry.
  To avoid cluttering, we shall not write all the terms of $V_{C\Phi H}$ but one should notice that terms such as $\lambda_C|H|^2|C|^2$ and $\lambda_{C'}|H|^2|C'|^2$ open up the possibility of the Higgs decaying into millicharged particles which would appear as $H\to invisibles$. From the bound on $Br(H\to invisibles)$ \cite{pdg}, we conclude $\lambda_C,\lambda_{C'}\stackrel{<}{\sim}0.02.$
 
 Let us now break the global $U(1)_D$ symmetry with the following Yukawa  term
 $$ Y_e \bar{e} \Phi_D^\dagger L_e+H.c.$$
 Via the tree level diagram shown in Fig \ref{decay}, we obtain 
 \be \frac{1}{\Lambda_C}= \frac{Y_e \lambda_{CC'}}{m_{\Phi^0}^2}\frac{v}{\sqrt{2}}. \ee
 Notice that the effective Lagrangian in Eq. (\ref{cc'}) breaks both the global $U_D(1)$ symmetry and the $Z_2$ symmetry under which $C' \to -C'$ so in the limit that $Y_e \lambda_{CC'}$ is  zero, the coupling of Eq. (\ref{cc'}) should vanish. In other words, for vanishing $Y_e$ or $\lambda_{CC'}$, the effective term in Eq. (\ref{cc'}) cannot be obtained at any loop level.
 The components of $\Phi_D$, having electroweak interaction, cannot be very light. The strongest lower bounds on their masses still come from LEP and are around 100 GeV \cite{pdg}.
 Notice that because in our model, $\Phi_D$ does not couple to quarks its only production mode is electroweak (vector fusion and associated production along with gauge boson \cite{Hashemi}). That is why the LHC cannot still 
 compete with LEP. On the other hand from unitarity consideration, strong upper bounds of 700-800 GeV are set on the masses of these particles \cite{Gorczyca:2011he}. We therefore expect $m_{\Phi^0}$ to be of order of  a few 100 GeV. Taking into account these bounds, we find
 \be Y_e \lambda_{CC'} \sim 10^{-13} \frac{10^{16} {\rm GeV}}{\Lambda_C}\left(\frac{500~{\rm GeV}}{m_{\Phi^0}}\right)^2. \label{num}\ee
The smallnesses of $Y_e$ and $\lambda_{CC'}$ are explained by the approximate $U(1)_D$ and the approximate $Z_2$ symmetry, respectively.
 
 The upper bounds on the masses of the $\Phi_D$ components guarantee their eventual discovery at the 
 high luminosity LHC. For negative (positive) $\lambda_2$,  the charged component of $\Phi_D$, $\Phi^+$, will be heavier (lighter) than the neutral component of $\Phi_D$, $\Phi^0$.
 Notice that as long as $\lambda_1+\lambda_2+\sqrt{\lambda \lambda_D}>0$ (where $\lambda$ is the SM Higgs quartic coupling), the ``unbounded from below" constraint will be satisfied even for negative $\lambda_2$ \cite{Gorczyca:2011he}.
 Let us discuss the case of the heavier $\Phi^+$ first and then discuss the case that $\Phi^+$ is lighter than 
 $\Phi^0$. If $\Phi^+$ is heavier than $\Phi^0$, the rate of $\Phi^+ \to \Phi^0 (W^+)^*$ (where $(W^+)^*$ is either on-shell or off-shell) can dominate over that of $\Phi^+ \to \nu e^+$. 
  If $Y_e \stackrel{>}{\sim} \lambda_{CC'} v/m_{\Phi^0}$, the $\Phi^0$ particle will  dominantly decay into $e^-e^+$ pair.  In the opposite case ({\it i.e.,} when  $Y_e \stackrel{<}{\sim} \lambda_{CC'} v/m_{\Phi^0}$), the $\Phi^0$ particle will mainly decay into the  $C'\bar{C}$ pair which appears as missing energy at detector. Since we do not want to open up a new production mode for the $C$ and $C'$ particle in the early universe which may affect the CMB and 21 cm line measurements, let us assume $\lambda_e \gg \lambda_{CC'} v/m_{\Phi^0}$. 
  This assumption along with the relation in Eq. (\ref{num}) implies $\Phi^0$ will immediately decay into the $e^-e^+$ pair with a lifetime shorter than $6.6\times 10^{-13}$ sec, so its signature will be a pair of 
 $e^-e^+$ with invariant mass corresponding to $m_{\Phi^0}$.
 Thus, to discover $\Phi^0$, the high luminosity mode of the LHC may focus on the gauge associated production of $\Phi^0$ which consists of a pair of $e^-e^+$ with a definite invariant mass corresponding to $m_{\Phi^0}$ and a SM gauge boson. This signal should be accompanied by a gauge associated production of $\Phi^+$ and its subsequent decay into $\Phi^0$ and $(W^+)^*$. Thus, the signature will be an $e^-e^+$ pair with invariant mass again equal to
 $m_{\Phi^0}$ and an on-shell or an off-shell $W$ boson plus an additional SM gauge boson.

 In the opposite case that $\Phi^+$ is lighter than $\Phi^0$, its main decay mode will be into $e^+ \nu_e$ pair.
 Thus the signature of the $\Phi^+$ production will be a SM gauge boson accompaned by a positron plus missing energy. In this case, the $\Phi^0$ ($\bar{\Phi}^0$) particle decays into $\Phi^+$ ($\Phi^-$) and $W^{-*}$
 ($W^{+*}$). The $\Phi^0$ ($\bar{\Phi}^0$) production in association with a gauge boson will lead into the signature of $e^+$ ($e^-$)
 plus missing energy along with a SM gauge boson.

 In our model, the new doublet couples exclusively to the leptons of the first generation. As a result,
 the decay of $\Phi^0$ and $\Phi^+$ produce only the first generation of the leptons. Moreover, the $C$ decay produces only the $e^-e^+$  flux, accounting for the AMS-02 signal. We could couple $\Phi_D$ to other fermions, in particular to the first generation of quarks. Then, $\Phi^0$ and $\Phi^+$ decays at colliders could produce quarks, appearing as pairs of jets. Moreover, the $C$ decay in the galaxy could produce quarks  which might contribute to the recently reported antiproton excess by AMS-02 \cite{Aguilar:2016kjl} but exploring
 this possibility is beyond the scope of the present paper.

\section{Summary and discussion} \label{Summary}

We have proposed a dark matter decay model solution to the positron excess observed by PAMELA and AMS-02. Within our model, dark matter, $X$,
 is a meta-stable particle which decays into  a pair of millicharged particles, $C\bar{C}$. If decay takes place in a region like  Milky Way galactic disk  where the background magnetic  field is high, the produced millicharged particles can be trapped despite the fact that their speed exceeds the gravitational escape velocity.
The $C$ and $\bar{C}$ particles eventually go through three body decay into $e^-e^+$ pair plus lighter millicharged particle. At production,  $e^-$ or $ e^+$  will have an energy between $m_C^2/(2 m_X)$ and $m_X/2$; however, they will lose energy because of interaction with interstellar matter and synchrotron radiation  before
reaching the detector.

Taking into account this energy loss and the uncertainties in the standard prediction for positron component of cosmic ray, we have surveyed the model parameter space to find the best fit to the positron excess observed by AMS-02. We have  found that the exact best fit point value depends on the  assumption on the positron energy loss function (see table \ref{table})  but overally with $m_X=1-10$ TeV, $m_C=1-10$ GeV and $\Gamma_X=10^{-27}-10^{-26}~{\rm sec}^{-1}$ a remarkable fit with a $p$-value above 70 \% can be found.
We also check for the compatibility of the predictions of our model with the $e^-+e^+$ spectrum measured by AMS-02 and CALET and found a reasonable goodness of fit.
 Thus, within our model, the entire positron excess can be explained by dark matter decay and there is no need for any extra contribution from pulsars or supernova remnants. 
If future studies establish pulsars and supernova remnants as powerful contributors to this excess, the AMS-02 data can be used to set a lower bound on $\Gamma_X$.
As mentioned before, within our model, we do not expect any significant gamma ray from dark matter halo. However, an isotropic gamma ray signal is expected from cumulation of
the photons produced by interaction of $e^{\pm}$ off CMB all over the universe. Dedicated analysis of the Fermi-LAT data and its successors must be carried out to account
for this effect. 

As described in \cite{Feldman:2007wj}, the millicharged particles can obtain their charge by adding a $U(1)$ gauge symmetry to the electroweak gauge group with a gauge boson
that mixes with the  hypercharge gauge boson. We denote the new gauge boson with $\gamma'$. We have also described how the effective coupling required for $C\to C' e^-e^+$ can be embedded in a viable electroweak invariant model. We have discussed the distinct predictions of this model for the high luminosity LHC.

We have discussed the possible bounds from various terrestrial, astrophysical and cosmological observations. The  $C'\bar{C'}$  pairs (as well as $C\bar{C}$)   can be produced and thermalized with the plasma in the early universe via the Drell-Yann mechanism and contribute as a millicharged component to dark matter on which there are strong bounds
from direct dark matter search experiments \cite{Dunsky:2018mqs}.  To prevent this,  a new annihilation mode is required to render the density  of millicharged relics small enough.
 The annihilation can lead to the production of new gauge bosons: $C\bar{C} \to \gamma'\gamma', C'\bar{C'} \to \gamma'\gamma'$. Efficient 
annihilation points towards light $\gamma'$ as well as light $C$ and $C'$ which opens up the prospect to test the model with terrestrial experiments. $\gamma'$ can be searched for by beam dump experiments such as SHiP \cite{Graverini:2015dka} and $C$ and $C'$ can be searched for with a setup such as SLAC millicharged experiment.
Even with maximal $\sigma(C\bar{C} \to \gamma'\gamma')$ and $\sigma(C'\bar{C'} \to \gamma'\gamma')$ (within the perturbative regime), $C'$ (produced either directly or via $C$ decay) can compose up to
$10^{-7}-10^{-8}$ of dark matter. Considering that the charges of $C$ and $C'$  should be about $10^{-6} $ within our model, the relic $C'$ can be eventually detected by direct dark matter search experiments. In fact for the parameter range of our interest, the bounds from direct dark matter search experiments on the fraction of the millichared particles are close to this limit \cite{Dunsky:2018mqs} so our model seems to be super-testable.

\section*{Acknowledgments}

This project has received funding from the European Union's Horizon 2020 research and innovation programme
under the Marie Sklodowska-Curie grant agreement No 674896 and No 690575. 
   M.R.  is grateful to Instituto de Fisica Teorica (IFT UAM-CSIC) in Madrid, where a part of this work was completed, for the hospitality of its staff and  for its support via the Centro de Excelencia Severo Ochoa Program under Grant SEV-2016-0597. M.R. would like to thank Pouya Bakhti and Alireza Vafaei Sadr for useful comments. Y.F. acknowledges ICTP associate office for the partial financial support.


\begin{thebibliography}{99}
\bibitem{Nagano:2000ve}
  M.~Nagano and A.~A.~Watson,
  Rev.\ Mod.\ Phys.\  {\bf 72} (2000) 689.
  doi:10.1103/RevModPhys.72.689



\bibitem{Adriani:2008zr}
  O.~Adriani {\it et al.} [PAMELA Collaboration],
  Nature {\bf 458} (2009) 607
  doi:10.1038/nature07942
  [arXiv:0810.4995 [astro-ph]].


\bibitem{Barwick:1997ig}
  S.~W.~Barwick {\it et al.} [HEAT Collaboration],
  Astrophys.\ J.\  {\bf 482} (1997) L191
  doi:10.1086/310706
  [astro-ph/9703192].










\bibitem{Aguilar:2013qda}
  M.~Aguilar {\it et al.} [AMS Collaboration],
  Phys.\ Rev.\ Lett.\  {\bf 110} (2013) 141102.
  doi:10.1103/PhysRevLett.110.141102




\bibitem{Hooper:2008kg}
  D.~Hooper, P.~Blasi and P.~D.~Serpico,
  JCAP {\bf 0901} (2009) 025
  doi:10.1088/1475-7516/2009/01/025
  [arXiv:0810.1527 [astro-ph]].


\bibitem{DiMauro:2014iia}
M.~Di Mauro, F.~Donato, N.~Fornengo, R.~Lineros and A.~Vittino,
JCAP {\bf 1404} (2014) 006
doi:10.1088/1475-7516/2014/04/006
[arXiv:1402.0321 [astro-ph.HE]].

\bibitem{Fujita:2009wk}
Y.~Fujita, K.~Kohri, R.~Yamazaki and K.~Ioka,
Phys.\ Rev.\ D {\bf 80} (2009) 063003
doi:10.1103/PhysRevD.80.063003
[arXiv:0903.5298 [astro-ph.HE]].

\bibitem{Kohri:2015mga}
K.~Kohri, K.~Ioka, Y.~Fujita and R.~Yamazaki,
PTEP {\bf 2016} (2016) no.2,  021E01
doi:10.1093/ptep/ptv193
[arXiv:1505.01236 [astro-ph.HE]].



\bibitem{Dunsky:2018mqs}
  D.~Dunsky, L.~J.~Hall and K.~Harigaya,
  arXiv:1812.11116 [astro-ph.HE].




\bibitem{ArkaniHamed:2008qn}
  N.~Arkani-Hamed, D.~P.~Finkbeiner, T.~R.~Slatyer and N.~Weiner,
  Phys.\ Rev.\ D {\bf 79} (2009) 015014
  doi:10.1103/PhysRevD.79.015014
  [arXiv:0810.0713 [hep-ph]].


\bibitem{Belotsky:2014nba}
  K.~Belotsky, M.~Khlopov and M.~Laletin,
  Bled Workshops Phys.\  {\bf 15} (2014) no.2,  1
  [arXiv:1411.3657 [hep-ph]].




\bibitem{Belotsky:2014haa}
  K.~Belotsky, M.~Khlopov, C.~Kouvaris and M.~Laletin,
  Adv.\ High Energy Phys.\  {\bf 2014} (2014) 214258
  doi:10.1155/2014/214258
  [arXiv:1403.1212 [astro-ph.CO]].
\bibitem{Hooper:2017gtd}
  D.~Hooper, I.~Cholis, T.~Linden and K.~Fang,
  Phys.\ Rev.\ D {\bf 96} (2017) no.10,  103013
  doi:10.1103/PhysRevD.96.103013
  [arXiv:1702.08436 [astro-ph.HE]].


\bibitem{Abeysekara:2017old}
  A.~U.~Abeysekara {\it et al.} [HAWC Collaboration],
  Science {\bf 358} (2017) no.6365,  911
  doi:10.1126/science.aan4880
  [arXiv:1711.06223 [astro-ph.HE]].


\bibitem{Yuan:2013eja}
  Q.~Yuan, X.~J.~Bi, G.~M.~Chen, Y.~Q.~Guo, S.~J.~Lin and X.~Zhang,
  Astropart.\ Phys.\  {\bf 60} (2015) 1
  doi:10.1016/j.astropartphys.2014.05.005
  [arXiv:1304.1482 [astro-ph.HE]].


\bibitem{Guo:2009aj}
  W.~L.~Guo and Y.~L.~Wu,
  Phys.\ Rev.\ D {\bf 79} (2009) 055012
  doi:10.1103/PhysRevD.79.055012
  [arXiv:0901.1450 [hep-ph]].







\bibitem{Slatyer:2009yq}
  T.~R.~Slatyer, N.~Padmanabhan and D.~P.~Finkbeiner,
  Phys.\ Rev.\ D {\bf 80} (2009) 043526
  doi:10.1103/PhysRevD.80.043526
  [arXiv:0906.1197 [astro-ph.CO]].




\bibitem{Liu:2018uzy}
  H.~Liu and T.~R.~Slatyer,
  Phys.\ Rev.\ D {\bf 98} (2018) no.2,  023501
  doi:10.1103/PhysRevD.98.023501
  [arXiv:1803.09739 [astro-ph.CO]].



\bibitem{Blanco:2018esa}
C.~Blanco and D.~Hooper,
arXiv:1811.05988 [astro-ph.HE].



\bibitem{Cirelli:2012ut}
M.~Cirelli, E.~Moulin, P.~Panci, P.~D.~Serpico and A.~Viana,
Phys.\ Rev.\ D {\bf 86} (2012) 083506
doi:10.1103/PhysRevD.86.083506, 10.1103/PhysRevD.86.109901
[arXiv:1205.5283 [astro-ph.CO]].

\bibitem{Kim:2017qaw}
  D.~Kim, J.~C.~Park and S.~Shin,
  JHEP {\bf 1804} (2018) 093
  doi:10.1007/JHEP04(2018)093
  [arXiv:1702.02944 [hep-ph]].



\bibitem{meshkat}
Y.~Farzan and M.~Rajaee,
JHEP {\bf 1712} (2017) 083
doi:10.1007/JHEP12(2017)083
[arXiv:1708.01137 [hep-ph]].


\bibitem{Sun:2010sm}
  X.~Sun and W.~Reich,
  Res.\ Astron.\ Astrophys.\  {\bf 10} (2010) 1287
  doi:10.1088/1674-4527/10/12/009
  [arXiv:1010.4394 [astro-ph.GA]].









\bibitem{Chuzhoy:2008zy}
L.~Chuzhoy and E.~W.~Kolb,
JCAP {\bf 0907} (2009) 014
doi:10.1088/1475-7516/2009/07/014
[arXiv:0809.0436 [astro-ph]].


\bibitem{Jansson:2012pc}
  R.~Jansson and G.~R.~Farrar,
  Astrophys.\ J.\  {\bf 757} (2012) 14
  doi:10.1088/0004-637X/757/1/14
  [arXiv:1204.3662 [astro-ph.GA]].



\bibitem{Benito:2019ngh}
M.~Benito, A.~Cuoco and F.~Iocco,
arXiv:1901.02460 [astro-ph.GA].



\bibitem{Cirelli:2010xx}
  M.~Cirelli {\it et al.},
  JCAP {\bf 1103} (2011) 051
   Erratum: [JCAP {\bf 1210} (2012) E01]
  doi:10.1088/1475-7516/2012/10/E01, 10.1088/1475-7516/2011/03/051
  [arXiv:1012.4515 [hep-ph]].


\bibitem{2016-AMS02-CERN}
 AMS-02 collaboration, The first five years of the Alpha Magnetic Spectrometer on the
International Space Station: unlocking the secrets of the cosmos, in AMS Five Years Data
Release, http://www.ams02.org/, 2016.





\bibitem{Tomassetti:2017hbe}
  N.~Tomassetti,
  Phys.\ Rev.\ D {\bf 96} (2017) no.10,  103005
  doi:10.1103/PhysRevD.96.103005
  [arXiv:1707.06917 [astro-ph.HE]].



\bibitem{Bindi:2017nee}
  V.~Bindi, C.~Corti, C.~Consolandi, J.~Hoffman and K.~Whitman,
  Adv.\ Space Res.\  {\bf 60} (2017) 865.
  doi:10.1016/j.asr.2017.05.025




\bibitem{Adriani:2018ktz}
  O.~Adriani {\it et al.},
  Phys.\ Rev.\ Lett.\  {\bf 120} (2018) no.26,  261102
  doi:10.1103/PhysRevLett.120.261102
  [arXiv:1806.09728 [astro-ph.HE]].


\bibitem{Feng:2012gs}
  L.~Feng, Q.~Yuan, X.~Li and Y.~Z.~Fan,
  Phys.\ Lett.\ B {\bf 720} (2013) 1
  doi:10.1016/j.physletb.2013.01.060
  [arXiv:1206.4758 [astro-ph.HE]].
\bibitem{Ambrosi:2017wek}
G.~Ambrosi {\it et al.} [DAMPE Collaboration],
Nature {\bf 552} (2017) 63
doi:10.1038/nature24475
[arXiv:1711.10981 [astro-ph.HE]].



\bibitem{Loparco:2017spm}
  F.~Loparco [Fermi-LAT Collaboration],
  J.\ Phys.\ Conf.\ Ser.\  {\bf 934} (2017) no.1,  012016.
  doi:10.1088/1742-6596/934/1/012016


\bibitem{Mambrini:2015sia}
  Y.~Mambrini, S.~Profumo and F.~S.~Queiroz,
  Phys.\ Lett.\ B {\bf 760} (2016) 807
  doi:10.1016/j.physletb.2016.07.076
  [arXiv:1508.06635 [hep-ph]].




\bibitem{Feldman:2008xs}
  D.~Feldman, Z.~Liu and P.~Nath,
  Phys.\ Rev.\ D {\bf 79} (2009) 063509
  doi:10.1103/PhysRevD.79.063509
  [arXiv:0810.5762 [hep-ph]].


\bibitem{Galli:2013dna}
S.~Galli, T.~R.~Slatyer, M.~Valdes and F.~Iocco,
Phys.\ Rev.\ D {\bf 88} (2013) 063502
doi:10.1103/PhysRevD.88.063502
[arXiv:1306.0563 [astro-ph.CO]].

\bibitem{Foot:2014uba}
R.~Foot and S.~Vagnozzi,
Phys.\ Rev.\ D {\bf 91} (2015) 023512
doi:10.1103/PhysRevD.91.023512
[arXiv:1409.7174 [hep-ph]].
\bibitem{Feldman:2007wj}
D.~Feldman, Z.~Liu and P.~Nath,
Phys.\ Rev.\ D {\bf 75} (2007) 115001
doi:10.1103/PhysRevD.75.115001
[hep-ph/0702123 [HEP-PH]].


\bibitem{Graverini:2015dka}
  E.~Graverini {\it et al.} [SHiP Collaboration],
  JINST {\bf 10} (2015) no.07,  C07007
  doi:10.1088/1748-0221/10/07/C07007
  [arXiv:1503.08624 [hep-ex]].

\bibitem{Fradette:2014sza}
  A.~Fradette, M.~Pospelov, J.~Pradler and A.~Ritz,
  Phys.\ Rev.\ D {\bf 90} (2014) no.3,  035022
  doi:10.1103/PhysRevD.90.035022
  [arXiv:1407.0993 [hep-ph]].




\bibitem{Batell:2009di}
  B.~Batell, M.~Pospelov and A.~Ritz,
  Phys.\ Rev.\ D {\bf 80} (2009) 095024
  doi:10.1103/PhysRevD.80.095024
  [arXiv:0906.5614 [hep-ph]].


\bibitem{Izaguirre:2013uxa}
  E.~Izaguirre, G.~Krnjaic, P.~Schuster and N.~Toro,
  Phys.\ Rev.\ D {\bf 88} (2013) 114015
  doi:10.1103/PhysRevD.88.114015
  [arXiv:1307.6554 [hep-ph]].
  
  
  
  
\bibitem{Batell:2014mga}
  B.~Batell, R.~Essig and Z.~Surujon,
  Phys.\ Rev.\ Lett.\  {\bf 113} (2014) no.17,  171802
  doi:10.1103/PhysRevLett.113.171802
  [arXiv:1406.2698 [hep-ph]].
  
\bibitem{Kahn:2014sra}
  Y.~Kahn, G.~Krnjaic, J.~Thaler and M.~Toups,
  Phys.\ Rev.\ D {\bf 91} (2015) no.5,  055006
  doi:10.1103/PhysRevD.91.055006
  [arXiv:1411.1055 [hep-ph]].
  

\bibitem{Bonivento:2013jag}
  W.~Bonivento {\it et al.},
  arXiv:1310.1762 [hep-ex].





\bibitem{Deshpande:1977rw}
  N.~G.~Deshpande and E.~Ma,
  Phys.\ Rev.\ D {\bf 18} (1978) 2574.
  doi:10.1103/PhysRevD.18.2574



\bibitem{Barbieri:2006dq}
  R.~Barbieri, L.~J.~Hall and V.~S.~Rychkov,
  Phys.\ Rev.\ D {\bf 74} (2006) 015007
  doi:10.1103/PhysRevD.74.015007
  [hep-ph/0603188].




\bibitem{LopezHonorez:2006gr}
  L.~Lopez Honorez, E.~Nezri, J.~F.~Oliver and M.~H.~G.~Tytgat,
  JCAP {\bf 0702} (2007) 028
  doi:10.1088/1475-7516/2007/02/028
  [hep-ph/0612275].




  \bibitem{pdg}
M. Tanabashi et al. (Particle Data Group), Phys. Rev. D 98, 030001 (2018).
\bibitem{Gorczyca:2011he}
B.~Gorczyca and M.~Krawczyk,
arXiv:1112.5086 [hep-ph].

\bibitem{Hashemi}
Y.~Farzan and M.~Hashemi,
JHEP {\bf 1011} (2010) 029
doi:10.1007/JHEP11(2010)029
[arXiv:1009.0829 [hep-ph]].

\bibitem{Prinz:1998ua}
  A.~A.~Prinz {\it et al.},
  Phys.\ Rev.\ Lett.\  {\bf 81} (1998) 1175
  doi:10.1103/PhysRevLett.81.1175
  [hep-ex/9804008].
\bibitem{Boehm:2013jpa}
  C.~Boehm, M.~J.~Dolan and C.~McCabe,
  JCAP {\bf 1308} (2013) 041
  doi:10.1088/1475-7516/2013/08/041
  [arXiv:1303.6270 [hep-ph]].


\bibitem{Chang:2018rso}
  J.~H.~Chang, R.~Essig and S.~D.~McDermott,
  JHEP {\bf 1809} (2018) 051
  doi:10.1007/JHEP09(2018)051
  [arXiv:1803.00993 [hep-ph]].
\bibitem{Aguilar:2016kjl}
M.~Aguilar {\it et al.} [AMS Collaboration],
Phys.\ Rev.\ Lett.\  {\bf 117} (2016) no.9,  091103.


\end{thebibliography}
\end{document}